\documentclass{JHEP3}

\newcommand{\dN}{\mathbb N}

\newcommand{\dS}{\mathbb S}
\newcommand{\dZ}{\mathbb Z}

\newcommand{\tm}{\tilde{\mu}}
\newcommand{\tn}{\tilde{\nu}}

\title{Dirac quantization of membrane winding uniformly \\in time dependent orbifold}
\author{Przemys{\l}aw Ma{\l}kiewicz$^\dag$ and W{\l}odzimierz Piechocki$^\ddag$
\\ Theoretical Physics Department\\Institute for Nuclear Studies
\\ Ho\.{z}a 69, 00-681 Warszawa, Poland;
\\ E-mail: $^\dag$pmalk@fuw.edu.pl, $^\ddag$piech@fuw.edu.pl}

\received{\today}       %%
\preprint{\arXivid{0905.3900}}

\abstract{We present quantum theory of a membrane propagating in
the vicinity of a time dependent orbifold singularity. The
dynamics of a membrane, with the parameters space topology of a
torus, winding uniformly around compact dimension of the embedding
spacetime is mathematically equivalent to the dynamics of a closed
string in a flat FRW spacetime. The construction of the physical
Hilbert space of a membrane makes use of the kernel space of
self-adjoint constraint operators. It is a subspace of the
representation space of the constraints algebra. There exist
non-trivial quantum states of a membrane evolving across the
singularity.}


\begin{document}

\section{Introduction}

In the cyclic universe scenario
\cite{Steinhardt:2001vw,Steinhardt:2001st} an evolution of the
universe consists of the sequence of classical and quantum phases.
One examines the possibility of describing each of quantum phases
in terms of quantum elementary objects in higher dimensional
($d>4$) compactified Milne, CM, space. The CM space includes a
cosmological singularity which consists of big-crunch and big-bang
epochs \cite{Khoury:2001bz}. Propagation of elementary objects
across the singularity is the main concern.

A reasonable model  of spacetime with the cosmological singularity
should allow for propagation of a quantum p-brane (i.e., particle,
string, membrane,...) from the pre-singularity to the
post-singularity epoch. If the quantum $p$-brane cannot go through
the cosmological  singularity, the cyclic evolution cannot be
realized. In our previous papers  we have examined the evolution
of a particle \cite{Malkiewicz:2005ii,Malkiewicz:2006wq} and a
string \cite{Malkiewicz:2006bw,Malkiewicz:2008dw} across the
singularity. A model of the quantum phase in the above sense seems
to be well defined.

The case of a membrane is technically more complicated because
functions describing membrane dynamics depend on three variables.
The Hamilton equations for these functions constitute a system of
coupled non-linear equations in higher dimensional phase space.
The Cauchy problem for membranes has been examined so far only in
ambient Lorentz {\it manifolds} \cite{Milbredt:2008sf}. Owing to
this complexity, we only try to identify  some {\it non-trivial}
membrane states  which propagate through the cosmological
singularity.

The first-class constraints describing membrane dynamics  are
generators of gauge transformations in the phase space of the
system and come from the reparametrization invariance of an action
integral. The goal of the present paper is the construction of a
quantum theory of a membrane winding around compact dimension of
CM space. The Hilbert space of a quantum membrane is constructed
by making use of the kernel space of the constraints
\cite{PAM,HT}.

Motivation for the present work is strictly cosmological, in spite
of the fact that the model has been inspired by string/M theory
\cite{Khoury:2001bz}. We do not intend to address in this paper
any problem of the M theory. Getting inside into the cosmological
problem is our main concern. However, our quantization seems to be
a novelty in the field of strings and it might be extended beyond
considered cosmological model.

In this paper we consequently use the Dirac method for quantizing
the Hamiltonian systems with constraints, which in principle may
be applied to the quantization of any $p$-brane propagating in
{\it curved} spacetime. Our method is different from the one used
in the textbooks on string theory (see, e.g. \cite{JP}). This is
probably why the critical dimension of spacetime does not occur in
our results.

The paper is organized as follows: Sec II deals with essentials of
the classical formulation of the system. In Sec III we define an
algebra of Hamiltonian constraints of a membrane. In Sec IV we
interpret constraints as generators for coordinate
transformations. Sec V concerns finding the representation of the
constraints algebra. We consider a toy model to get some insight
into the problem by using a single field. Then, we use some fields
defined on the phase space of the membrane in the background space
to construct the physical Hilbert space.  We conclude in the last
section.

\section{Hamiltonian}

The Polyakov action for a test $p$-brane   embedded in a
background spacetime with  metric $g_{\tm\tn}$ has the form
\begin{equation}\label{act}
    S_p= -\frac{1}{2}\mu_p \int d^{p+1}\sigma
    \sqrt{-\gamma}\;\big(\gamma^{ab}\partial_a X^{\tm} \partial_b
    X^{\tn}
    g_{\tm\tn}-(p-1)\big),
\end{equation}
where $\mu_p$ is a mass per unit $p$-volume,
$(\sigma^a)\equiv(\sigma^0,\sigma^1,\ldots,\sigma^p)$ are
$p$-brane worldvolume coordinates, $\gamma_{ab}$ is the $p$-brane
worldvolume metric, $\gamma := det[\gamma_{ab}]$,
$~(X^{\tm})\equiv (X^\mu, \Theta)\equiv (T,X^k,\Theta)\equiv
(T,X^1,\ldots,X^{d-1},\Theta)$ are the embedding functions of a
$p$-brane, i.e. $X^{\tm} = X^{\tm}(\sigma^0,\ldots,\sigma^p$), in
$d+1$ dimensional background spacetime.

It has been found  \cite{Turok:2004gb} that the total Hamiltonian,
$H_T$, corresponding to the action (\ref{act}) is the following
\begin{equation}\label{ham}
H_T = \int d^p\sigma \mathcal{H}_T,~~~~\mathcal{H}_T := A C + A^i
C_i,~~~~~i=1,\ldots,p
\end{equation}
where $A=A(\sigma^a)$ and $A^i = A^i(\sigma^a)$ are any functions
of $p$-volume coordinates,
\begin{equation}\label{conC}
    C:=\Pi_{\tm} \Pi_{\tn} g^{\tm\tn} + \mu_p^2 \;det[\partial_a X^{\tm} \partial_b
    X^{\tn} g_{\tm\tn}]\approx 0,
\end{equation}
\begin{equation}\label{conCi}
    C_i := \partial_i X^{\tm} \Pi_{\tm} \approx 0,
\end{equation}
where $\Pi_{\tm}$ are the canonical momenta corresponding to
$X^{\tm}$,  and where the symbol `$\approx $' denotes `weakly
zero' in the sense of Dirac \cite{PAM}. Equations (\ref{conC}) and
(\ref{conCi}) define the first-class constraints of the system.

The Hamilton equations are
\begin{equation}\label{hameq}
    \dot{X}^{\tm}\equiv\frac{\partial{X}^{\tm}}{\partial\tau}=
    \{X^{\tm},H_T \},~~~~~~\dot{\Pi}_{\tm}\equiv\frac{\partial{\Pi}_{\tm}}{\partial\tau}=
    \{\Pi_{\tm},H_T \},~~~~~~\tau\equiv\sigma^0,
\end{equation}
where the Poisson bracket is defined by
\begin{equation}\label{pois}
    \{\cdot,\cdot\}:= \int d^p\sigma\Big(\frac{\partial\cdot}{\partial X^{\tm}}
    \frac{\partial\cdot}{\partial \Pi_{\tm}}
     - \frac{\partial\cdot}{\partial \Pi_{\tm}}
    \frac{\partial\cdot}{\partial X^{\tm}}\Big).
\end{equation}

In what follows we restrict our considerations to the compactified
Milne, CM, space. The CM space is one of the simplest models of
spacetime implied by string/M theory \cite{Khoury:2001bz}. Its
metric is defined by the line element
\begin{equation}\label{line}
    ds^2 = -dt^2 +dx^k dx_k + t^2 d\theta^2 = \eta_{\mu\nu} dx^\mu
    dx^\nu  + t^2 d\theta^2 = g_{\tm\tn} dx^{\tm} dx^{\tn},
\end{equation}
where $\eta_{\mu\nu}$ is the Minkowski metric, and $\theta$
parameterizes   a circle. Orbifolding $\dS^1$ to a segment
$~\dS^1/\dZ_2~$  gives the model of spacetime in the form of two
planes  which collide and re-emerge at $t=0$. Such model of
spacetime has been used in
\cite{Steinhardt:2001vw,Steinhardt:2001st}. Our results do not
depend on the choice of topology of the compact dimension.

In  our previous paper \cite{Malkiewicz:2006bw} and in the present
one we analyze the dynamics of a $p$-brane  which is {\it winding
uniformly} around the $\theta$-dimension. For the $p$-brane in
such a state, we identify the $p$-th worldvolume coordinate with
the compact dimension $\theta$ and impose the condition that the
state of the $p$-brane does not depend on this coordinate, i.e.
\begin{equation}\label{con1}
    \sigma^p = \theta =\Theta~~~~~~\mbox{and}~~~~~\partial_\theta X^\mu =0=\partial_\theta
    \Pi_\mu,
\end{equation}
which leads to
\begin{equation}\label{con2}
    \frac{\partial}{\partial\theta}(X^{\tm})=
    (0,\ldots,0,1)~~~~~\mbox{and}~~~~~\frac{\partial}{\partial\tau}(X^{\tm})=
    (\dot{T},\dot{X}^k,0).
\end{equation}
The conditions (\ref{con1}) reduce  (\ref{conC})-(\ref{pois}) to
the form in which the canonical pair $(\theta,\Pi_\theta)$ does
not occur \cite{Turok:2004gb}.

\section{Algebra of constraints}

In the case of a winding uniformly membrane applying the condition
(\ref{con1}) to the formulae (\ref{conC}) and (\ref{conCi}) leads
to the two  constraints
\begin{equation}\label{cond2}
C = \Pi_\mu(\tau,\sigma)\;\Pi_\nu (\tau,\sigma)\;\eta^{\mu\nu}
    + \kappa^2\;T^2(\tau,\sigma) \acute{X}^\mu (\tau,\sigma)
    \acute{X}^\nu (\tau,\sigma)\;\eta_{\mu\nu}\approx 0,
\end{equation}
\begin{equation}\label{cond3}
    C_1= \acute{X}^\mu (\tau,\sigma)\;\Pi_\mu(\tau,\sigma)\approx 0,~~~~~C_2 = 0,
\end{equation}
where $\acute{X}^\mu := \partial X^\mu/\partial \sigma~,~~\sigma
:=\sigma^1,$ $\kappa := \theta_0\mu_2$,  and where $\theta_0:=\int
d\theta$.

To examine the algebra of constraints we `smear' the constraints
as follows
\begin{equation}\label{integ}
    \check{A}:= \int_{-\pi}^\pi d\sigma \;
    f(\sigma)A(X^{\mu}, \Pi_{\mu}),~~~~f \in \{C^\infty [-\pi,\pi]\,|\,
    f^{(n)}(-\pi)=f^{(n)}(\pi)\}.
\end{equation}
The Lie bracket is defined as
\begin{equation}\label{alg}
\{\check{A},\check{B}\}:= \int_{-\pi}^\pi d\sigma\;
\Big(\frac{\partial\check{A}}{\partial X^\mu}
    \frac{\partial\check{B}}{\partial \Pi_\mu}
     - \frac{\partial\check{A}}{\partial \Pi_\mu}
    \frac{\partial\check{B}}{\partial X^\mu}\Big).
\end{equation}
The constraints in an integral form satisfy the  algebra
\begin{equation}\label{aalg1}
\{\check{C}(f_1),\check{C}(f_2)\}=  \check{C}_1 \big(4\kappa^2 T^2
(f_1 \acute{f}_2 - \acute{f}_1 f_2)\big),
\end{equation}
\begin{equation}\label{aalg2}
\{\check{C}_1(f_1),\check{C}_1(f_2)\}= \check{C}_1(f_1 \acute{f}_2
- \acute{f}_1 f_2),
\end{equation}
\begin{equation}\label{aalg3}
\{\check{C}(f_1),\check{C}_1(f_2)\}= \check{C}(f_1 \acute{f}_2 -
\acute{f}_1 f_2).
\end{equation}
Equations (\ref{aalg1})-(\ref{aalg3}) demonstrate that $C$ and
$C_1$ are first-class constraints because the Poisson algebra
closes. However, it is not  a Lie algebra because the factor $T^2$
is not a constant, but a function on phase space. Little is known
about representations of such type of an algebra. Similar
mathematical problem occurs in  general relativity (see, e.g.
\cite{TT}).

The smearing (\ref{integ}) of constraints  helps to get the
closure of the algebra in an explicit form. A local form of the
algebra includes the Dirac delta so the algebra makes sense but in
the space of distributions (see Appendix A for more details). It
seems that such an arena is inconvenient for finding a
representation of the algebra which is required in the
quantization procedure.

The original algebra of constraints may be rewritten in a
tractable form by making use of the redefinitions
\begin{equation}\label{AAA}
C_{\pm}:=\frac{C\pm C_1}{2}
\end{equation}
where
\begin{equation}\label{BBB}
C := \frac{\mbox{\scriptsize{original}}~C}{ 2\kappa T},~~~~C_1 :=
\mbox{\scriptsize{original}}~C_1,
\end{equation}
where `original' means defined by    (\ref{cond2}) and
(\ref{cond3}). The new algebra reads
\begin{equation}\label{alg1}
    \{ \check{C}_{+}(f),
    \check{C}_{+}(g)\} = \check{C}_{+}(f\acute{g}-g\acute{f}),
\end{equation}
\begin{equation}\label{alg2}
    \{ \check{C}_{-}(f),
    \check{C}_{-}(g)\} = \check{C}_{-}(f\acute{g}-g\acute{f}),
\end{equation}
\begin{equation}\label{alg3}
\{ \check{C}_{+}(f), \check{C}_{-}(g)\} = 0 .
\end{equation}
The redefined algebra is a Lie algebra.

The redefinition (\ref{BBB}) seems to be a technical trick without
a physical interpretation. In what follows we show that it
corresponds to the specification of the winding zero-mode state of
a membrane not at the level of the phase space, but at the level
of an action integral.

The Nambu-Goto action for a membrane in the CM space reads
\begin{eqnarray}
    S_{NG}&=&-\mu_2\int d^3\sigma\sqrt{-det(\partial_aX^{\mu}\partial_bX^{\nu}
    g_{\mu\nu})}\\ \label{membrane}
    &=&-\mu_2\int
d^3\sigma\sqrt{-det(-\partial_aT\partial_bT+T^2\partial_a\Theta\partial_b
\Theta+\partial_aX^k\partial_bX_k )}
\end{eqnarray}
where $(T,\Theta ,X^k)$ are embedding functions of the membrane
corresponding to the spacetime coordinates $(t,\theta,x^k)$
respectively.

An action  $S_{NG}$ in the lowest energy winding mode,  defined by
(\ref{con2}), has the form
\begin{eqnarray}\label{string}
    S_{NG}&=&-\mu_2\theta_0\int
d^2\sigma\sqrt{-T^2det(-\partial_aT\partial_bT+\partial_aX^k\partial_bX_k
)}\\ \label{string2} &=&-\mu_2\theta_0\int
d^2\sigma\sqrt{-det(\partial_aX^{\alpha}\partial_bX^{\beta}\widetilde{g}_{\alpha\beta})}.
\end{eqnarray}
where $a,b\in \{0,1\}$ and
$\widetilde{g}_{\alpha\beta}=T\eta_{\alpha\beta}$.  Thus, the
propagation of a membrane in this special mode is equivalent to
the evolution of a string in the spacetime with dimension $d$
(while $d+1$ was the original one), which is now  a flat
Friedmann-Robertson-Walker spacetime, $ds^2=t\eta_{\alpha\beta}$.

One can verify that the Hamiltonian corresponding to the string
action (\ref{string2}) has the form
\begin{equation}\label{hamiltonian}
H_T = \int d\sigma \mathcal{H}_T,~~~~\mathcal{H}_T := A C + A^1
C_1,
\end{equation}
where
\begin{equation}\label{con}
    C:=\frac{1}{2\mu_2\theta_0  T}\Pi_{\alpha} \Pi_{\beta}\eta^{\alpha\beta} +
    \frac{\mu_2\theta_0}{2} \;T\;\partial_a X^{\alpha} \partial_b
    X^{\beta} \eta_{\alpha\beta}\approx 0,~~~~
    C_1 := \partial_{\sigma} X^{\alpha} \Pi_{\alpha} \approx 0,
\end{equation}
and  $A=A(\tau,\sigma)$ and $A^1 = A^1(\tau,\sigma)$ are any
regular functions. Therefore (\ref{con}) and (\ref{BBB}) coincide,
which gives an interpretation for the redefinition of the
constraints.

\section{Meaning of the constraints}

An action integral of a string is invariant with respect to smooth
and invertible maps of worldsheet coordinates
\begin{equation}\label{iso}
    (\tau, \sigma)\rightarrow (\tau', \sigma').
\end{equation}
These diffeomorphisms  considered infinitesimally form an algebra
of local fields $-\epsilon(\tau, \sigma)\partial_{\tau}$ and
$-\eta(\tau, \sigma)\partial_{\sigma}$ (we refer to their actions
on the fields as $\dot{\delta}_{\epsilon}$ and $\delta'_{\eta}$,
respectively). Mapping (\ref{iso}) leads to the infinitesimal
changes of the fields $X^{\mu}(\tau, \sigma)$ and $\Pi_{\mu}(\tau,
\sigma)=\partial L/\partial \dot{X}^{\mu}=\mu(\frac{1}{A}g_{\mu
\nu}\dot{X}^{\nu}-\frac{A^1}{A}g_{\mu \nu}\acute{X}^{\nu})$ as
follows
\begin{equation}\label{iso2}
    \delta X^{\mu}=\dot{\delta}_{\epsilon}X^{\mu}+\delta'_{\eta}X^{\mu}=
    \epsilon\dot{X}^{\mu}+\eta\acute{X}^{\mu},~~~~~
    \delta\Pi_{\mu}=\epsilon\dot{\Pi}_{\mu}+\acute{\epsilon}
    (A^1{\Pi}_{\mu}+\mu Ag_{\mu
    \nu}\acute{X}^{\nu})+(\eta{\Pi}_{\mu})'.
\end{equation}
The transformations (\ref{iso2}) are defined along curves in the
phase space with coordinates $(X^{\mu},\Pi_{\mu})$ and are
expected to be generated by the first-class constraints
$\check{C}$ and $\check{C}_1$ according to the theory of gauge
systems \cite{PAM,HT}. One verifies  that
\begin{equation}\label{dif1}
\{X^{\mu}, \check{C}(\varphi)\}=\frac{\varphi}{\mu}\Pi_{\mu},~~~~
\{\Pi_{\mu},
\check{C}(\varphi)\}=-\frac{\varphi}{2\mu}(\Pi_{\alpha}
 \Pi_{\beta}g^{\alpha\beta}_{,X^{\mu}}+\acute{X}^{\alpha}
 \acute{X}^{\beta}g_{\alpha\beta ,X^{\mu}})
 +\mu(\varphi g_{\mu \nu}\acute{X}^{\nu})',
\end{equation}
\begin{equation}\label{dif4}
\{X^{\mu}, \check{C}_1(\phi)\}=\phi\acute{X}^{\mu},~~~~
\{\Pi_{\mu},\check{C}_1(\phi)\}=(\phi\Pi_{\mu})',
\end{equation}
where $\phi(\sigma, \tau)$ and $\varphi(\sigma, \tau)$ are
smearing functions depending on two variables, and the integration
defining the smearing of the constraints $C$ and $C_1$ does not
include the integration with respect to $\tau$ variable (see
(\ref{integ})).

The comparison  of (\ref{iso2}) with (\ref{dif1})-(\ref{dif4})
gives specific relations between these two transformations. For
the action of the constraints along curves in the phase space,
which are solutions to the equations of motion, we get
\begin{equation}\label{diff1}
\{X^{\mu}, \check{C}(\varphi)\}=
\dot{\delta}_{\frac{\varphi}{A}}X^{\mu}-\delta'_{\frac{A^1\varphi}{A}}X^{\mu},
~~~~\{\Pi_{\mu},
\check{C}(\varphi)\}=\dot{\delta}_{\frac{\varphi}{A}}\Pi_{\mu}-\delta'_{\frac{A^1\varphi}
{A}}\Pi_{\mu},
\end{equation}
\begin{equation}\label{diff4}
\{X^{\mu}, \check{C}_1(\phi)\}=\delta'_{\phi}X^{\mu},~~~~
\{\Pi_{\mu}, \check{C}_1(\phi)\}=\delta'_{\phi}\Pi_{\mu}.
\end{equation}
We see that the relation between gauge transformations in the
string coordinate space and its phase space may be established
only for curves (coordinate transformations are not
time-invariant), which are solutions to the Hamilton equations.
This is why the above relation depends on the specific choice of
$A$ and $A^1$.

\section{Representations}

This section is devoted to the Dirac quantization of the system.
It consists of two essential steps: (i) definition of a
self-adjoint representation of the algebra of constraints on a
\emph{kinematical} Hilbert space, and (ii) solution to the
constraints, i.e. finding the common domain on which \emph{all}
the constraint operators vanish, which is used to construct a
\emph{physical} Hilbert space.

It is clear that (\ref{alg1})-(\ref{alg3}) consists of two
independent subalgebras. To be specific, we first quantize the
subalgebra satisfied by
\begin{equation}\label{cir}
    L_n := \check{C}_{+}(\exp{in\sigma}),~~~~n \in \dZ.
\end{equation}

The imposition of the constraint (\ref{cir}) with $n \in \dZ$ (but
not with $n \in \dN$, that is specific to the conformal field
theory quantization) is consistent since for the real fields
$X^{\mu}$ and $\Pi_{\mu}$ holds $\overline{L_n} = L_{-n}$ and the
constraints $L_n$ and $L_{-n}$ are equivalent. Moreover, including
the complex conjugation for the classical constraints allows for
introducing adjoint constraint operators at quantum level as we
shall see later.

One may easily verify that
\begin{equation}\label{alg4}
\{L_n,L_m\} = i (m-n) L_{m+n}.
\end{equation}

Quantization of (\ref{alg2}) can be done by analogy.  Merger of
both quantum subalgebras will complete the problem of finding the
representation of the full algebra (\ref{alg1})-(\ref{alg3}).  To
construct the representation of the algebra
(\ref{alg1})-(\ref{alg3}), which consists of two commuting
subalgebras, one may use standard techniques
\cite{EP,Malkiewicz:2006bw}. For instance, the representation
space of the algebra may be defined to be either a tensor product
or direct sum of the representations of both subalgebras.

\subsection{Representation based on a  single field}

\subsubsection{Hilbert space}

The pre-Hilbert space, $\tilde{\mathcal{H}}$, induced by the space
of fields, $\dS \ni \sigma \rightarrow X(\sigma)$, is defined to
be
\begin{eqnarray}\label{state}
  \tilde{\mathcal{H}}\ni\Psi[X] &:=& \int \psi(X,\acute{X},\sigma)d\sigma ,
  \\\label{scal1}
  \langle\Psi |\Phi \rangle &:=& {\int\overline{\Psi}[X]\Phi[X][dX]},
\end{eqnarray}
where $\psi(X,\acute{X},\sigma)$ is  such that $ \langle\Psi |\Psi
\rangle<\infty$. The measure $[dX]$ is assumed to be invariant
with respect to $\sigma$ reparametrization. Completion of
$\tilde{\mathcal{H}}$ in the norm induced by (\ref{scal1}) defines
the Hilbert space $\mathcal{H}$.

\subsubsection{Representation of  generator}

In what follows we find a  representation of (\ref{alg4}). Let us
consider a diffeomorphism on $\dS^1$ of the form $X(\sigma)
\mapsto X(\sigma+\epsilon v(\sigma))$. For a small $\epsilon$ we
have
\begin{eqnarray}\label{xx1}
  X(\sigma+\epsilon
v(\sigma)) &\approx& X(\sigma)+\epsilon v(\sigma)\acute{X}
(\sigma)=: X(\sigma)+\epsilon L_{v}X(\sigma), \\
  \acute{X}(\sigma+\epsilon
v(\sigma))&\approx&
\acute{X}(\sigma)+\epsilon\frac{d}{d\sigma}[v(\sigma)\acute{X}(\sigma)]
= \acute{X}(\sigma)+\epsilon\frac{d}{d\sigma}[L_{v}{X}(\sigma)].
\end{eqnarray}
Now, we define an operator $\hat{L}_{v}$ corresponding to  $L_{v}$
defined by (\ref{xx1}). Since we have
\begin{equation}
\Psi[X(\sigma+\epsilon
v(\sigma))]\approx\Psi[X(\sigma)]+\epsilon\int\Big(
\frac{\partial\psi}{\partial X}L_{v}X+\frac{\partial\psi}{\partial
\acute{X}}\frac{d}{d\sigma}[L_{v}{X}]\Big)d\sigma,
\end{equation}
we set
\begin{equation}\label{l}
\hat{L}_{v}\Psi[X]:=\int \Big(\frac{\partial\psi}{\partial
X}L_{v}X+\frac{\partial\psi}{\partial
\acute{X}}\frac{d}{d\sigma}[L_{v}{X}]\Big)d\sigma=\int
\Big(\acute{v}\frac{\partial\psi}{\partial
\acute{X}}\acute{X}-\acute{v}\psi-v\frac{\partial\psi}{\partial
\sigma}\Big) d\sigma~\in\mathcal{H}.
\end{equation}
One may verify that $\{L_v,L_w \}=L_{(v\acute{w}-\acute{v}w)}$ and
check that
\begin{equation}\label{aalg}
[\hat{L}_{v},\hat{L}_{w}]=\hat{L}_{(v\acute{w} -\acute{v}w)}.
\end{equation}
Next, let us consider the following
\begin{eqnarray}\nonumber
\int\overline{\Psi}[X(\sigma+\epsilon
v(\sigma))]\Phi[X(\sigma)][dX(\sigma)]&=&\int\overline{\Psi}[X(\sigma)]
\Phi[X(\sigma-\epsilon v(\sigma))][dX(\sigma-\epsilon
v(\sigma))]\\ \label{conj}
&=&\int\overline{\Psi}[X(\sigma)]\Phi[X(\sigma-\epsilon
v(\sigma))][dX(\sigma)],
\end{eqnarray}
where we assume that $v(\sigma)$ is a real function and
$\sigma\mapsto \sigma+\epsilon v(\sigma)$ is a diffeomorphism.
Taking derivative with respect to $\epsilon$ of both sides of
(\ref{conj})   and putting $\epsilon=0$ leads to
\begin{equation}\label{algN}
\langle\hat{L}_v\Psi |\Phi \rangle=-\langle\Psi |\hat{L}_v\Phi
\rangle .
\end{equation}
Therefore, the real and imaginary parts of the operator $\hat{L}_n
$ defined by the mapping
\begin{equation}\label{Map1}
    L_n \mapsto \hat{L}_n :=  i\hbar \,\hat{L}_{\exp(i n\sigma)}
\end{equation}
are symmetric on $\mathcal{H}$ and lead to a symmetric
representation of the algebra (\ref{alg4}), which now reads
\begin{equation}\label{quantum}
\frac{1}{i\hbar}\,[\hat{L}_n,\hat{L}_m] = i(m-n) \hat{L}_{m+n}
\end{equation}
The equation (\ref{quantum}) is a self-adjoint representation of
(\ref{alg4}) if $\hat{L}_n$ are bounded operators \cite{MRS}.

\subsubsection{Solving the constraint}

Since we look for diffeomorphism invariant states,  it is
sufficient to assume that $\psi=\psi(X,\acute{X})$. Let us solve
the equation
\begin{equation}\label{sss}
\hat{L}_n\Psi=0,
\end{equation}
which after making use of (\ref{l}) and integrating by parts reads
\begin{equation}\label{cons1}
\int \acute{(e^{\imath
n\sigma})}[-\psi+\frac{\partial\psi}{\partial
\acute{X}}\acute{X}]~d\sigma=0.
\end{equation}
General solution to (\ref{cons1}) has the form
\begin{equation}\label{cons}
-\psi+\frac{\partial\psi}{\partial \acute{X}}\acute{X}=\sum_{k\neq
-n}a_k e^{\imath k\sigma}~~~~\textrm{for}~~~n\neq 0 ,
\end{equation}
where $a_k$ are arbitrary constants, and there is no condition for
$n=0$. Our goal is an imposition of all the constraint,  i.e. we
look for  $\Psi :\forall n~\hat{L}_n\Psi=0$. We find that the
intersection of all the kernels defined by (\ref{cons}) is given
by the equation
\begin{equation}\label{conss}
-\psi+\frac{\partial\psi}{\partial \acute{X}}\acute{X}=c,
\end{equation}
where $c$ is an arbitrary constant. It is enough to solve
(\ref{conss})  for $c=0$ and then simply add to the solution any
constant. Since the above equation results from (\ref{cons1}), it
is expected to hold in a more general sense, i.e. in a
distributional sense. It is clear that the space of solutions to
(\ref{conss}) is defined by
\begin{equation}\label{ssol}
\psi=\alpha(X)|\acute{X}|+\beta(X)\acute{X}-c,
\end{equation}
where $\alpha$ and $\beta$ are any functions. The first term is a
distribution, the second one can be checked to be trivial, since
\begin{equation}
\int_{\dS^1} \beta(X)\acute{X}~d\sigma=\int_{\dS^1} \beta(X) dX=0
\end{equation}
for a periodic field $X$, and third one is a functional that gives
the same value $2\pi c$ for every field.

Notice that taking $n>0$ (as in the conformal field theory) would
reduce (\ref{cons}) to the form
\begin{equation}\label{eqq}
-\psi+\frac{\partial\psi}{\partial \acute{X}}\acute{X}=\sum_{k\neq
-n}a_k e^{\imath k\sigma}~~~~\textrm{for}~~~n> 0 ,
\end{equation}
and subsequently would lead to the solution of (\ref{eqq}) in the
form
\begin{equation}\label{sssol}
\psi=\alpha(X)|\acute{X}|+\beta(X)\acute{X}-\sum_{k>-1}a_k
e^{\imath k\sigma},
\end{equation}
instead of (\ref{conss}), which is not diffeomorphism invariant
due to the last term.   The diffeomorphism is the basic symmetry
underlying our paper.

\subsubsection{Interpretation of solutions}

Let us identify special features of the fields $X$ evaluated
through the first term in (\ref{sssol})
\begin{eqnarray}\nonumber
\Psi[X] &=& \int \alpha(X)|\acute{X}|~d\sigma = \int
\frac{d}{d\sigma}{[\gamma(X)]}(\tilde{H}(\acute{X})-\tilde{H}(-\acute{X}))~d\sigma
 \\ \label{dist}
&=& -\int \gamma(X)2\delta(\acute{X})~d\acute{X} =-
\sum_{\textrm{extr}\;  X }2\gamma(X)= \sum_{\textrm{min}\; X
}2\gamma(X)-\sum_{ \textrm{max}\; X }2\gamma(X),
\end{eqnarray}
where $d\gamma/dX=\alpha$ and $\tilde{H}$ is the Heaviside
function. Thus, $\Psi$  depends on the values of $\gamma$ at
extrema points of $X$. We have diffeomorphism invariance due to
the implication $(\frac{dX}{d\sigma}=0)\Rightarrow
(\frac{dX}{d\widetilde{\sigma}}=
\frac{d\sigma}{d\widetilde{\sigma}}\frac{dX}{d\sigma}=0)$.

\subsection{Representation based on phase space functions}

\subsubsection{Hilbert space}

Using the ideas with  the single field case (presented in the
previous subsection), we try construct now the representation of
the algebra (\ref{alg1})-(\ref{alg3}) by making use of the phase
space functions with coordinates $(X^{\mu}, \Pi_{\mu})$, where
$\mu=0,1,\dots, d-1$.

 As before, we propose to include fields $~\dS \ni \sigma
\rightarrow Y^{\mu}(\sigma)~$ as well as their first derivatives
$\acute{Y}^{\mu}(\sigma)$ in the definition of a state
\begin{eqnarray}\label{multistate}
  \mathcal{H}\ni\Psi[\overrightarrow{Y}] &:=& \int \psi(\overrightarrow{Y},
  \acute{\overrightarrow{Y}},\sigma)d\sigma ,\\
  \langle\Psi |\Phi \rangle &:=& \int\overline{\Psi}[\overrightarrow{Y}]
  \Phi[\overrightarrow{Y}][d\overrightarrow{Y}],
\end{eqnarray}
where $\overrightarrow{Y} \equiv (Y^\mu)$, and where
$\psi(\overrightarrow{Y},\acute{\overrightarrow{Y}},\sigma)$ is
any well-behaved function such that $ \langle\Psi |\Psi
\rangle<\infty$.

Inspired by \cite{Thiemann:2004qu},  we try to find two sets of
phase space functions $Y_{\pm}$ by the requirement that they
Poisson commute with the constraints:
\begin{eqnarray}\label{main}
    0 = \{C_{\pm}(u),Y_{\pm}(v)\}&=&\int -\bigg[\frac{d}{d\sigma}\bigg(\frac{u}{2}
    \kappa T \acute{X}_{\nu}\bigg)\pm
    \frac{d}{d\sigma}\bigg(\frac{u}{2}\Pi_{\nu}\bigg)\bigg]\frac{\delta Y_{\pm}
    (v)}{\delta
    \Pi_{\nu}}\\ \nonumber  &+&\bigg[-\frac{u}{4}\frac{\Pi^2}{\kappa
    T^2}+\frac{u}{4}\kappa \acute{X}^2\bigg]\frac{\delta Y_{\pm}(v)}{\delta
    \Pi_{0}}-\bigg[\frac{u}{2}\frac{\Pi^{\nu}}{\kappa T}\pm\frac{u}{2}\acute{X}^{\nu}
    \bigg]\frac{\delta Y_{\pm}(v)}{\delta
    X^{\nu}}~d\sigma .
\end{eqnarray}
The indices are lowered or raised by $\eta_{\mu\nu}$ or
$\eta^{\mu\nu}$, respectively, and
\begin{equation}\label{var}
\frac{\delta Y_{\pm}(v)}{\delta
    Z}:=\Sigma_{n=0}^{\infty}(-1)^n\bigg(\frac{\partial Y_{\pm}(v)}{\partial
    Z^{(n)}}v\bigg)^{(n)} ,
\end{equation}
where $\big(\ldots \big)^{(n)}$ denotes the n-th derivative with
respect to $\sigma$.

Finding solution to Eq. (\ref{main}) determines two domains of
definition of quantum states. Imposition of a constraint on each
of them defines two physical phase spaces.   Hilbert space of the
whole system may be defined, for instance, as a direct sum or a
tensor product of these two Hilbert spaces.

Let us introduce new coefficients
\begin{equation}
D_{\pm}^{\nu} :=\frac{1}{2}\Pi^{\nu}\pm \frac{1}{2}\kappa T
\acute{X}^{\nu}
%B &=& \frac{1}{4}\frac{\Pi^2}{\kappa
%    T^2}-\frac{1}{4}\kappa \acute{X}^2
\end{equation}
so that (\ref{main}), after taking into account (\ref{var}), reads
\begin{eqnarray}\nonumber
\{C_{\pm}(u),Y_{\pm}(v)\}&=&\int v\Sigma_{n=0}^{\infty} \bigg[\mp
(uD_{\pm}^{\nu})^{(n+1)}\frac{\partial Y_{\pm}}{\partial
    (\Pi^{\nu})^{(n)}}+\bigg(u\frac{D_{+}^{\mu}D_{-\mu}}{\kappa T^2}\bigg)^{(n)}
    \frac{\partial Y_{\pm}}{\partial
    (\Pi^{0})^{(n)}}\\
&-&\bigg(u\frac{D_{\pm}^{\nu}}{\kappa T}\bigg)^{(n)}\frac{\partial
Y_{\pm}}{\partial
    (X^{\nu})^{(n)}}\bigg]~d\sigma .
\end{eqnarray}
The condition $\{C_{\pm}(u),Y_{\pm}(v)\}=0$ must hold for any $v$,
hence
\begin{equation}
\Sigma_{n=0}^{\infty}\bigg[ \mp
(uD_{\pm}^{\nu})^{(n+1)}\frac{\partial Y_{\pm}} {\partial
    (\Pi^{\nu})^{(n)}}+\bigg(u\frac{D_{+}^{\mu}D_{-\mu}}{\kappa T^2}\bigg)^{(n)}
    \frac{\partial Y_{\pm}}{\partial
    (\Pi^{0})^{(n)}}-\bigg(u\frac{D_{\pm}^{\nu}}{\kappa T}\bigg)^{(n)}\frac{\partial
    Y_{\pm}}{\partial
    (X^{\nu})^{(n)}}\bigg]=0  .
\end{equation}
As $u$ is arbitrary too, we obtain the following infinite set of
equations:
\begin{eqnarray}\nonumber
\Sigma_{n=m}^{\infty}\frac{n!}{(n-m)!}\bigg[ \mp
(D_{\pm}^{\nu})^{(n-m)}\frac{\partial Y_{\pm}}{\partial
(\Pi^{\nu})^{(n-1)}}+\bigg(\frac{D_{+}^{\mu}D_{-\mu}}{\kappa
T^2}\bigg)^{(n-m)} \frac{\partial Y_{\pm}}{\partial
    (\Pi^{0})^{(n)}}\\ \label{final}-\bigg(\frac{D_{\pm}^{\nu}}{\kappa T}\bigg)^{(n-m)}
    \frac{\partial Y_{\pm}}{\partial
    (X^{\nu})^{(n)}}\bigg]=0 ,
\end{eqnarray}
where $m=0,1 \dots$.

It is very difficult to solve the system (\ref{final}) in its full
generality.  Let us assume for simplicity  that $Y_{\pm} =
\tilde{Y}_{\pm} := Y_{\pm}(X,\acute{X},\Pi)$, i.e. we ignore
possible dependance on higher derivatives of $X$ and $\Pi$ with
respect to $\sigma$. In such a case an infinite system of
equations (\ref{final}) simplifies to the following system of only
two equations:
\begin{eqnarray}\label{e1}
    D^{\nu}_{\pm}\frac{\partial \tilde{Y}}{\partial D^{\nu}_{\pm}} &=&
    0,\\ \nonumber
   \mp 2\acute{D}^{\nu}_{\pm}\frac{\partial \tilde{Y}}{\partial D^{\nu}_{\pm}}+\frac{1}{\kappa
T^2}D_{-\nu}D_+^{\nu}\bigg(\frac{\partial \tilde{Y}}{\partial
    D_{+}^0}+\frac{\partial \tilde{Y}}{\partial
    D_{-}^0}\bigg)
    -2D_{\pm}^{\nu}\frac{1}{\kappa T}\frac{\partial \tilde{Y}}{\partial
    X^{\nu}}\\ -D_{\pm}^{0}\frac{1}{\kappa
T^2}\bigg(D_+^{\nu}-D_-^{\nu}\bigg)\bigg(\frac{\partial
\tilde{Y}}{\partial
    D_+^{\nu}}-\frac{\partial \tilde{Y}}{\partial
    D_-^{\nu}}\bigg) \mp \frac{1}{
    \kappa T^2}(D^0_+-D^0_-)D_{\pm}^{\nu}\frac{\partial \tilde{Y}}{\partial
    D_{\mp}^{\nu}}&=& 0 . \label{e2}
\end{eqnarray}
Due to the assumption that $\tilde{Y}_{\pm}$ does not depend
neither on $\acute{D}^{\nu}_{+}$ nor on $\acute{D}^{\nu}_{-}$, we
obtain that the solution to (\ref{e1}) reads $\tilde{Y}_{\pm} =
\tilde{Y}_{\pm}(X^{\mu},D^{\mu}_{\mp})$. Thus, Eq. (\ref{e2})
turns into
\begin{eqnarray}\nonumber
   \frac{1}{\kappa
T^2}D_{-\nu}D_+^{\nu}\frac{\partial \tilde{Y}_{\pm}}{\partial
    D_{\mp}^0}
    -2D_{\pm}^{\nu}\frac{1}{\kappa T}\frac{\partial \tilde{Y}_{\pm}}{\partial
    X^{\nu}}\noindent \\ \label{e3} \pm D_{\pm}^{0}\frac{1}{\kappa
T^2}\bigg(D_+^{\nu}-D_-^{\nu}\bigg)\frac{\partial
\tilde{Y}_{\pm}}{\partial
    D_{\mp}^{\nu}} \mp \frac{1}{
    \kappa T^2}(D^0_+-D^0_-)D_{\pm}^{\nu}\frac{\partial \tilde{Y}_{\pm}}{\partial
    D_{\mp}^{\nu}}&=& 0  .
\end{eqnarray}
Since $\tilde{Y}_{\pm}$ does not depend on $D^{\nu}_{\mp}$, we
conclude that Eq. (\ref{e3}) splits into the following system of
equations:
\begin{eqnarray}\label{e4}
    D^i_{\mp}\frac{\partial \tilde{Y}_{\pm}}{\partial
    D_{\mp}^{0}}-2T\frac{\partial \tilde{Y}_{\pm}}{\partial
    X^{i}}+D^0_{\mp}\frac{\partial \tilde{Y}_{\pm}}{\partial
    D_{\mp}^{i}}&=&0,~~~~\textrm{for}~~~~i=1,\dots,d\\ \label{e5}
    -2T\frac{\partial \tilde{Y}_{\pm}}{\partial
    T}-D^{\nu}_{\mp}\frac{\partial \tilde{Y}_{\pm}}{\partial
    D_{\mp}^{\nu}}&=&0 .
\end{eqnarray}
Equations  (\ref{e4}) and (\ref{e5}) have only two independent
solutions:
\begin{eqnarray}
\tilde{Y}_{\pm}&=& \frac{D_{\mp}^{\mu}D_{\mp\mu}}{\kappa
T}=C_{\mp}.
\end{eqnarray}
Such solutions have been expected due to
$\{C_{+}(u),C_{-}(v)\}=0$, for any $u$ and $v$. We have shown that
these are the only solutions to (\ref{main}) under the assumption
that $Y_{\pm}=Y_{\pm}(X,\acute{X},\Pi)$. More general solutions
may be found by admitting  that $Y_{\pm}$ depend on higher
derivatives of $X$ and $\Pi$. Considering of such generalizations
is, however,  beyond the scope of the present paper.

\subsubsection{Solving the constraint}

Now, let us make use the  solution to Eq. (\ref{main}) for each
region separately. We assume that
$\psi=\psi(\overrightarrow{Y},\acute{\overrightarrow{Y}})$. Let us
solve the equation
\begin{equation}
\hat{L}_n\Psi[\overrightarrow{Y}]=0,
\end{equation}
which in the case of many fields is a simple extension of
(\ref{cons1}), and reads
\begin{equation}\label{consy}
\int \acute{(e^{\imath
n\sigma})}[-\psi+\frac{\partial\psi}{\partial
\acute{Y}^{\mu}}\acute{Y}^{\mu}]~d\sigma=0.
\end{equation}
By analogy to the single field case we infer that
\begin{equation}
-\psi+\frac{\partial\psi}{\partial
\acute{Y}^{\mu}}\acute{Y}^{\mu}=\sum_{k\neq -n}a_k e^{\imath
k\sigma}~~~~\textrm{for}~n\neq 0
\end{equation}
and again with no condition for $n=0$. Imposing all the
constraints leads to
\begin{equation}
-\psi+\frac{\partial\psi}{\partial
\acute{Y}^{\mu}}\acute{Y}^{\mu}=c .
\end{equation}
One can check that the solutions form a linear space and are of
the form
\begin{equation}\label{hilbert}
\psi=\bigg(\sum_i\alpha_i(\overrightarrow{Y})\prod_{\mu}|
\acute{Y}^{\mu}|^{\rho_i^{\mu}}\bigg)^{\frac{1}{\rho}}-c,
\end{equation}
where $\sum_{\mu}\rho_i^{\mu}=\rho$. This is an expected result
since  the measure
$\sqrt[\rho]{\prod_{\mu}|\acute{Y}^{\mu}|^{\rho^{\mu}}}d\sigma$ is
invariant with respect to $\sigma$-diffeomorphisms.

\subsubsection{Interpretation of solutions}

Suppose we have a space $V\ni\overrightarrow{Y}$ in which a closed
curve, $\sigma\mapsto Y^{\mu}(\sigma)$, is embedded.  Due to
(\ref{hilbert}) we have a kind of measure in $V$ given by
\begin{equation}\label{measure}
\sqrt[\rho]{\alpha(\overrightarrow{Y})\prod_{\mu}|dY^{\mu}
|^{\rho^{\mu}}}.
\end{equation}
One may say, it is a generalization of the  Riemannian type
metric, since for  $\rho^{\mu}_i =1$ and $\rho=2$ we have
\begin{equation}
\sqrt{g_{\mu\nu}dY^{\mu}dY^{\nu}},
\end{equation}
where $g_{\mu\nu}=g_{\mu\nu}(\overrightarrow{Y})$. In the case,
e.g., $Y^0$ is not a constant field,  (\ref{measure}) becomes
\begin{equation}\label{measure1}
\sqrt[\rho]{\alpha(\overrightarrow{Y})\prod_{\mu}|dY^{\mu}|^{\rho^{\mu}}}
=\sqrt[\rho]{\alpha (\overrightarrow{Y})\prod_{\mu\neq
0}\bigg|\frac{dY^{\mu}}{dY^0}\bigg|^{\rho^{\mu}}}|dY^0| =:
\widetilde {\alpha}(Y^0)|dY^0|.
\end{equation}
Thus, it is an extension of the single field metric defined by
(\ref{dist}), which may be rewritten as $\alpha(Y)|dY|$. In this
case however integration (\ref{measure1}) is performed in the
multidimensional space so $\widetilde{\alpha}(Y^0)$  depends on a
particular curve  (not just its end points). In fact, it is a
measure of relative variation of fields, i.e. quantity that is
both gauge-invariant and determines curve uniquely. Two simple
examples of wavefunction for two fields $Y_1$ and $Y_2$ are given
by
\begin{eqnarray}\label{ex1}
  \psi &=& \alpha(Y_1\pm Y_2)|\acute{Y_1}\pm \acute{Y_2}|, \\ \label{ex2}
  \psi &=& \alpha(Y_1 Y_2)|\acute{Y_1}Y_2+Y_1\acute{Y_2}|,
\end{eqnarray}
where in analogy to the single field case, (\ref{ex1}) and
(\ref{ex2}) `measure extrema points' for fields $Y_1\pm Y_2$ and
$Y_1 Y_2$, respectively.

It is clear that finding the representation of the complete
algebra (\ref{alg1})-(\ref{alg3}), may be carried out by analogy
to the single field case by using  standard techniques
\cite{EP,Malkiewicz:2006bw}. For instance, we may define
$\Psi[Y^{\mu}_{+},~Y^{\mu}_{-}]:=\Psi[Y^{\mu}_{+}]\otimes\Psi[Y^{\mu}_{-}]$.

\subsection{Remarks on representations of observables}

In the space of solutions to the constraints there are many types
of measures in the form (\ref{measure}) which may be used to
define a variety of physical Hilbert spaces and representations.
One may associate operators, in physical Hilbert space,  with
homeomorphisms $V\mapsto V$. The operators split the Hilbert space
into a set of invariant subspaces, each of which defines a
specific representation. Each subspace is connected with specific
measure and all other measures that are produced by
homeomorphisms. For example, the products of the action of
homeomorphism upon a metric (of Riemannian manifold) constitute
the space of all the metrics that are equivalent modulo a change
of coordinates and all other metrics that are reductions of the
initial metric.

Now, let us consider an infinitesimal homeomorphism,
$\widehat{O}_u : V \rightarrow V$, of the space $V$ along the
vector field $u=u^{\lambda}(\overrightarrow{Y})\,\partial/\partial
Y^{\lambda}$. In what follows we consider an example of
representation:

For the special form of (\ref{hilbert}) defined by
\begin{equation}\label{rrep1}
\psi :=
\alpha_{\mu}(\overrightarrow{Y})\acute{Y}^{\mu},~~~~\textrm{or}~~
\Psi[Y]=\int\alpha_{\mu}(\overrightarrow{Y})d{Y}^{\mu},
\end{equation}
we find that \cite{Trautman}
\begin{equation}\label{tra1}
   \frac{1}{i\hbar}\, \widehat{O}_u\bigg(\int\alpha_{\mu}d{Y}^{\mu}\bigg)=\int\big(
u^{\lambda}\alpha_{\mu,\lambda}+u^{\lambda}_{,\mu}\alpha_{\lambda}\big)d{Y}^{\mu}.
\end{equation}

One may verify that the operators $\widehat{O}_u$ and
$\widehat{O}_v$ associated with vector fields $u$ and $v$ satisfy
the algebra
\begin{equation}\label{repal}
\frac{1}{i\hbar}\,[ \widehat{O}_u,\widehat{O}_v
]=\widehat{O}_{[u,v]}
\end{equation}
The representation defined by (\ref{tra1}) and (\ref{repal}) is
self-adjoint if the operators are bounded.

\section{Conclusions}

The quantization problem  of a membrane embedded in a time
dependent orbifold is difficult. It has not been solved
satisfactory so far even for the case of the Minkowski target
space (see, eg. \cite{Smolin:1997ai}). Most proposals for quantum
theory of membranes are based on finding relationships between
very special membrane states  and string states (see, e.g.
\cite{Turok:2004gb,Horava:2008ih}). In this paper we have
considered states of membrane  winding uniformly around compact
dimension of the background space.

An action integral of a membrane  winding uniformly around compact
dimension of CM space depends on functions of two variables and
there are only two  constraints. The dynamics of a membrane, with
the parameters space topology $\dS^1\times\dS^1$, winding
uniformly around compact dimension of embedding spacetime is
mathematically equivalent to the dynamics of a closed string in a
a flat FRW spacetime.

General quantum theory of a string in a {\it curved} spacetime has
not been constructed yet. We have proposed a framework which
includes curved background and is based on Dirac's quantization of
the diffeomorphism symmetry. However, it is different from the
quantization program initiated in \cite{Thiemann:2004qu}.

The first-class constraints specifying the dynamics of a membrane
propagating in the compactified Milne space satisfy the algebra
which is a  Poisson algebra. Methods for finding a self-adjoint
representation of such type of an algebra are complicated
\cite{TT}. We overcome this difficulty by the reduction and
redefinition of the constraints algebra. Resulting algebra is a
Lie algebra which simplifies the problem of quantization of the
membrane dynamics.

We found an example of solution to Eq. (\ref{main}) which enables
defining the Hilbert space of the system. The imposition of the
constraints led to the physical states. It means that the
singularity of the CM space {\it is not} an insurmountable
obstacle for the construction of a quantum model of a  system
describing propagation of a membrane. There exist non-trivial
states of a membrane evolving across the singularity. Quantum
states of a membrane winding uniformly around compact dimension of
the CM space are examples of such non-trivial states. There may
exist the membrane states which cannot be quantized by our method.
We postpone an examination of this issue to our next papers.

The quantum membrane completes our preliminary tests of the
`transparency' of the cosmological singularity of the cyclic model
of the universe: there  exist non-trivial quantum states of
$p$-branes (particle, string, membrane) that can propagate from
the pre-singularity to the post-singularity epoch. Extended
objects may `cure' the disappearance of the compact dimension at
the singularity, which is specific to the time dependent
orbifolds. The propagation of a particle across the singularity
may be indeterministic, whereas higher dimensional objects seem to
propagate uniquely.  The big-crunch/big-bang model of the universe
with quantum objects propagating in classical spacetime may make
sense. However, this is not the end of the story. We have not
examined all possible states. There may exist quantum $p$-brane
states which may lead to problems at the singularity and should be
analyzed.

There is still another problem to be examined  in the context of
the cyclic model: We have considered so far the propagation of
{\it test} $p$-branes (i.e. objects which do not modify the
background space). The $p$-branes which are {\it physical} may
lead to the gravitational instability. It has been argued
\cite{Horowitz:2002mw} that due to this problem the big-crunch of
the cyclic model may collapse into a black hole which would end
the evolution of the universe. In such a case, the cyclic model
scenario would need to be modified to make sense. One may study
this problem by quantization of the entire system, i.e. $p$-brane
{\it and} the embedding  spacetime, by making use of the loop
quantum cosmology.  We have already made some preliminary steps to
examine this problem \cite{Dzierzak:2009ip,Malkiewicz:2009qv}.

\subsection{Acknowledgments}

This work has been supported by the Polish Ministry of Science and
Higher Education Grant  NN 202 0542 33.

\appendix

\section{Local form of the constraints algebra}

It has been found (see Appendix 2 of \cite{Turok:2004gb} ) that
the constraints (\ref{cond2}) and (\ref{cond3}) satisfy the
algebra
\begin{equation}\label{a1}
\{C(\sigma),C(\sigma^{\prime})\}=8\kappa^2 T^2 (\sigma)\;C_1
(\sigma)\frac{\partial}{\partial \sigma}\delta
(\sigma^{\prime}-\sigma)+ 4\kappa^2 \delta
(\sigma^{\prime}-\sigma)\frac{\partial}{\partial \sigma}\big(T^2
(\sigma)C_1(\sigma)\big),
\end{equation}
\begin{equation}\label{a2}
\{C(\sigma),C_1 (\sigma^{\prime})\}= 2 \;C
(\sigma)\frac{\partial}{\partial \sigma}\delta
(\sigma^{\prime}-\sigma) + \delta
(\sigma^{\prime}-\sigma)\frac{\partial}{\partial \sigma}
C(\sigma),
\end{equation}
\begin{equation}\label{a3}
\{C_1 (\sigma),C_1 (\sigma^{\prime})\}= 2 \;C_1
(\sigma)\frac{\partial}{\partial \sigma}\delta
(\sigma^{\prime}-\sigma) + \delta
(\sigma^{\prime}-\sigma)\frac{\partial}{\partial \sigma}
C_1(\sigma),
\end{equation}
where $\partial X^\mu (\sigma^{\prime})/\partial X^\nu (\sigma) =
\delta^\mu_\nu \delta(\sigma^{\prime}-\sigma) =  \partial \Pi_\nu
(\sigma^{\prime})/\partial \Pi_\mu (\sigma)$ (with other partial
derivatives being zero), and where the Poisson bracket is defined
to be
\begin{equation}\label{locP}
\{\cdot,\cdot\}:= \int_{-\pi}^\pi d\sigma\;
\Big(\frac{\partial\cdot}{\partial X^\mu}
    \frac{\partial\cdot}{\partial \Pi_\mu}
     - \frac{\partial\cdot}{\partial \Pi_\mu}
    \frac{\partial\cdot}{\partial X^\mu}\Big).
\end{equation}

\end{document}